\documentclass{pazhastl}
\usepackage{graphicx}
\begin{document}

\journalinfo{2009}{35}{00}{001}[000]

\title{Kinematics of the Outer Pseudorings and the Spiral
  Structure of the Galaxy}

\author{A.~M.~Mel'nik\address{1}\email{anna@sai.msu.ru}   and P.~Rautiainen\address{2}
  \addresstext{1}{ Sternberg Astronomical Institute, Moscow, Russia}
  \addresstext{2}{ Astronomy  Division, Department of Physical Sciences,  University of Oulu, Finland}
  }

\shortauthor{MEL'NIK AND RAUTIAINEN}

\shorttitle{KINEMATICS OF THE OUTER PSEUDORINGS}

\submitted{November 11, 2008; in final form, February 04, 2009}

\begin{abstract}
The kinematics of the outer rings and pseudorings is determined by
two processes: the resonance tuning and the gas outflow.  The
resonance kinematics is clearly observed in the pure rings, while
the kinematics of the gas outflow is manifested itself in the
pseudorings. The direction of systematical motions in the pure
rings depends on the position angle of a point with respect to the
bar major axis and on the class of the outer ring. The direction
of the radial and azimuthal components of the residual velocities
of young stars in the Perseus, Carina, and Sagittarius regions can
be explained by the presence of the outer pseudoring of class
$R_1R_2'$ in the Galaxy. We present models, which reproduce the
directions  and values of the residual velocities  of
OB-associations in the Perseus and Sagittarius regions, and also
model reproducing the directions of the residual velocities in the
Perseus, Sagittarius, and Carina regions. The kinematics of the
Sagittarius region accurately defines the solar position angle
with respect to the bar elongation, $\theta_b=45\pm5^\circ$.

\end{abstract}

\section{INTRODUCTION}

\subsection{The Galactic spiral structure}

\begin{figure*}
\resizebox{\hsize}{!}{\includegraphics{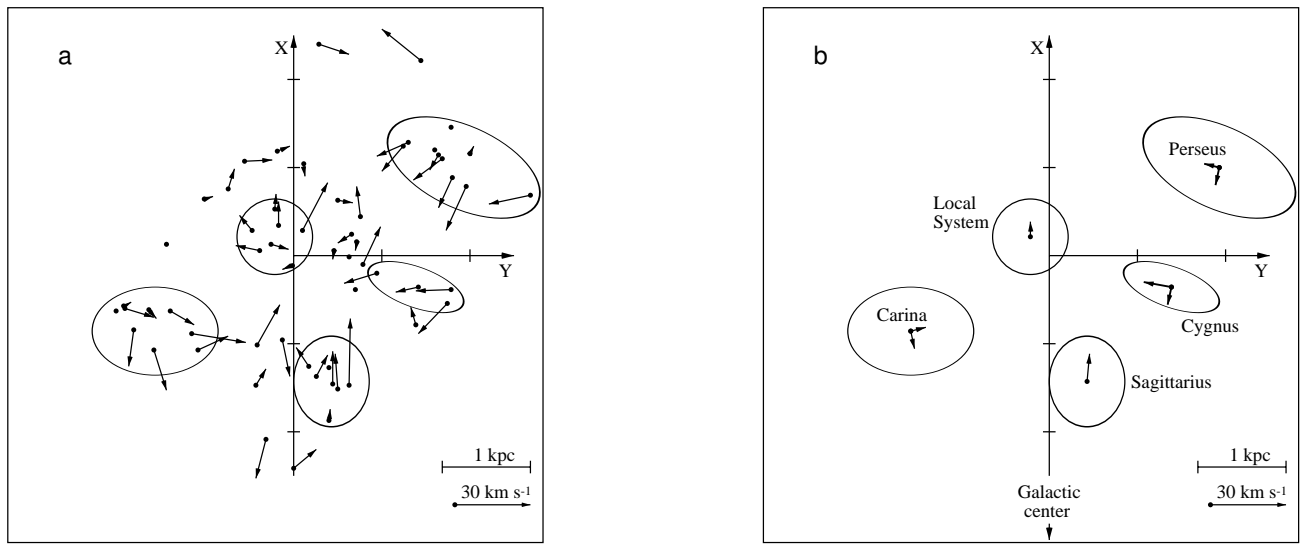}} \caption{ (a)
The residual velocities of OB-associations projected onto the
galactic plane. The union of OB-associations into regions of
intense star formation. (b) The average velocities, $V_R$ and
$V_\theta$,  of OB-associations in the regions of intense star
formation. The X-axis is directed away from the center, the Y-axis
is in the direction of the Galactic rotation. The Sun is at the
origin.}
\end{figure*}

The spiral structure is clearly observed in many external galaxies
viewed face-on, but in the Galaxy we are faced with a difficulty
of the distance determination for the indicators of spiral
structure. On  the other hand, in the Galaxy we can study the
field of space velocities using both line-of-sight velocities and
proper motions, whereas with other galaxies we are almost
completely limited to line-of-sight velocities.

One of the best tracers of the spiral structure are HII regions --
gas clouds which are ionized by young hot stars. These can be seen
as bright radio objects throughout the disc of the Galaxy
(Georgelin and Georgelin 1976;  Russeil 2003; and other papers).
Though the distances for the distant HII regions, $r>4$ kpc, are
determined from observations badly (usually they are derived from
the kinematical models) the distribution of HII regions displays
the most general features of the Galactic spiral structure which
can be formulated as follows.

\begin{itemize}

\item[1.] The Galactic spiral structure is most pronounced at the
  Galactocentric distances $R=5-9$ kpc.

\item[2.] The pitch   angle of the Galactic spirals is quite
small, $i<20^\circ$.

\item[3.] Many researchers believe the Galaxy is 4-armed.

\end{itemize}

In the vicinity of 3 kpc from the Sun the optical data has
revealed the existence of three fragments of the spiral structure:
Sagittarius-Carina, Cygnus-Orion, and Perseus ones. The
characteristic of these regions is the  intense star formation
manifested as increased concentration of young clusters and
OB-associations (Humphreys 1979; and other papers)

The Sagittarius-Carina and Perseus arm-fragments are often thought
to be part of the global spiral structure (Georgelin and Georgelin
1976; Efremov 1998; Russeil 2003; Vall\'ee 2005; and other
papers). The Cygnus-Orion fragment is usually regarded as the
local arm. This conception appears mostly because of its location
between two global arms.

The study of the Galactic structure in the neutral hydrogen
revealed the existence of regions with systematical non-circular
motions which were later identified with spiral arms (Burton
1971). The investigation of the Perseus region in OB-associations,
red supergiants,  molecular and neutral hydrogen showed the
presence of systematical motions which were interpreted as motions
directed towards the Galactic center (Burton and Bania 1974;
Humphreys 1976; Avedisova and Palous 1989; Brand and Blitz 1993;
and other papers). Recent studies agree  with this result (Melnik
et al. 2001; Sitnik 2003; and other papers).

The density-wave theory which had been already developed by the
end of the 60s (Lin et al. 1969, Lin 1970; Roberts 1969) afforded
an opportunity to determine the location of spiral arms with
respect to the corotation radius (CR) through the direction of gas
streaming motions in spiral arms. The direction of streaming
motions in the Perseus region strongly suggests that its location
is inside the CR.

The density-wave theory describes the kinematics of young stars
and gas in the Perseus, Cygnus, and Carina regions quite well
(Melnik 2003). Nevertheless its application to the whole 3-kpc
solar neighborhood encounters some difficulties. Young stars in
the Carina and Sagittarius regions, through which the spiral arm
is traditionally drawn, have different systematical motions and
cannot be fragments of the same density-wave spiral arm. Moreover,
the young stars with the systematical non-circular motions
directed towards the Galactic center (the Perseus, Cygnus, and
Carina regions) which could belong to the same, perhaps patchy,
spiral arm fall nicely on a large-scale structure which appears to
be the leading spiral arm (Melnik 2006).

In the present paper we intend to build the dynamical models that
reproduce the kinematics of young stars in the 3-kpc solar
neighborhood. We'll show that model of the Galaxy with an outer
ring of class $R_1R_2'$ can explain  the kinematics of young stars
in some regions.

\subsection[]{The kinematics of young stars within 3 kpc of the
Sun}

%--------- Table 1  ------------------
\begin{table*}
\small \caption {Average residual velocities of
OB-associations}\footnotesize
\begin{center}
 \begin{tabular}{lcccccl}
  \\[-7pt]\hline\\[-7pt]
   Region     & {\it R}, kpc & $V_R$, & $V_{\theta}$,  & {\it l}, deg. & {\it r}, kpc & Associations \\
    &  & km s$^{-1}$ & km s$^{-1}$  &  &  &  \\
  \\[-7pt]\hline\\[-7pt]
  Sagittarius & 5.6 & $+11\pm3$ & $-1\pm1$ & 8--23  & 1.3--1.9 & Sgr OB1, OB7, OB4, Ser OB1, OB2, \\
  & & & & & &  Sct OB2, OB3;\\
  Carina & 6.5 & $-6\pm2$ & $+5\pm3$ & 286--315  & 1.5--2.1 & Car OB1, OB2, Cru OB1, Cen OB1,\\
  & & & & & &   Coll 228, Tr 16, Hogg 16, NGC 3766, 5606;\\
  Cygnus & 6.9 & $-7\pm3$ & $-11\pm2$ & 73--78  & 1.0--1.8 & Cyg OB1, OB3, OB8, OB9; \\
  Local System & 7.4 & $+5\pm3$ & $+0\pm3$ & 0--360  & 0.1--0.6 & Per OB2, Mon OB1, Ori OB1, Vela OB2, \\
  & & & & & &   Coll 121, 140, Sco OB2; \\
  Perseus & 8.4 & $-7\pm2$ & $-5\pm2$ & 104--135  & 1.8--2.8 & Per OB1, NGC 457, Cas OB8, OB7, OB6, \\
  & & & & & &  OB5, OB4, OB2, OB1, Cep OB1;\\
 \hline
\end{tabular}
\end{center}
\end{table*}

OB-associations are the most suitable objects for the kinematical
investigations  in the wide  solar neighborhood. These loose
groups of high-luminosity stars have quite reliable distances (the
average accuracy is about 15\%). This good accuracy appears due to
young clusters which  often enter OB-associations (Garmany and
Stencel 1992). But unlike young clusters OB-associations contain a
sufficient number of stars with known line-of-sight velocities and
proper motions. On average the space velocities of OB-associations
are determined over 11 stars. The stellar proper motions were
taken from the Hipparcos catalog (1997). The electronic version of
the catalog of line-of-sight velocities and proper motions of
OB-associations is available at
http://lnfm1.sai.msu.ru/$\sim$anna/page3.html (for more details,
see Melnik et al. 2001).

Fig.~1a  shows the residual velocities of OB-associations in a
projection onto the Galactic plane. The residual velocities were
calculated as the differences between the observed heliocentric
velocities and the velocities due to the  circular rotation law
and the solar motion relative to the centroid of OB-associations.
The parameters of the circular Galactic rotation law and the
components of the solar motion were derived from the sample of
OB-associations located within 3 kpc from the Sun (Melnik et al.
2001). In such a large region systematical non-circular motions of
OB-associations can be regarded as random deviations from the
rotation curve. The obtained parameters describe the rotation of
the low-dispersed Galactic subsystem. Calculation of the residual
velocities with respect to the rotation curve derived from the
same objects yields minimal residual velocities, any other
rotation curve would produce larger, on average, deviations.
Besides, the use of the self-consistent distance scale also
decreases the deviations from the rotation curve (Sitnik and
Melnik 1996; Dambis et al. 2001). The derived rotation curve is
nearly flat and corresponds to the linear velocity at the solar
distance $\Theta_0=215$ km s$^{-1}$, the value of $\Theta$ remains
constant  inside 3 kpc from the Sun with the accuracy $\pm3$\%. We
adopted the Galactocentric distance of the Sun to be $R_0=7.1$ kpc
(Dambis at al. 1995; Glushkova et al. 1998; and other papers). The
distances for OB-associations from the catalog by Blaha and
Humphreys (1989) were multiplied by a factor of $0.8$ to be
consistent with the so-called short distance scale for classical
Cepheids (Berdnikov et al. 2000).

Fig.~1a also shows the union of OB-associations into regions of
intense star formation which practically coincide with the
stellar-gas complexes united by Efremov and Sitnik (1988). Fig.~1b
represents the average residual velocities of OB-associations in
regions of intense star formation in projection onto the radial,
$V_R$, and azimuthal, $V_\theta$, directions. It is a
generalization of Fig.~1a. The average residual velocities and
their random errors are  given in Table 1, which  also contains
the average Galactocentric distances $R$, the intervals of
galactic longitudes $l$, the intervals of heliocentric distances
$r$, and names of OB-association the region includes.

The study of the Hipparcos catalog (1997) shows that systematical
errors in positions of  bright stars ($m<9^m$) don't exceed 0.0001
arcsec (Kovalevsky, 2002). Stars of the  Blaha and Humphreys'
catalog (1989) having the known proper motion are bright enough,
their average magnitude equals $m_V=7.3^m$ in the region $0<r<3$
kpc and $m_V=7.8^m$ in the region $1.5<r<3$ kpc. Since the mission
of Hipparcos (1997) continued several years (37 months), we can
suppose that systematical errors of proper motions of stars of
OB-associations don't exceed 0.0001 arcsec yr$^{-1}$, that
corresponds to the velocity of 1 km s$^{-1}$ at the distance of
$r=2$ kpc. Besides, the contribution of systematical errors
decreases after averaging of proper motions over a large area:
most of OB-associations occupy on the sky more than 10 square
degrees.

The velocity $V_R$ in the Perseus, Cygnus and Carina regions is
directed towards the Galactic center and is about $V_R=-7$ km
s$^{-1}$. In the Sagittarius region it is directed away from the
Galactic center, $V_R=+11$ km s$^{-1}$. This means that
Sagittarius and Carina regions cannot belong to the same trailing
density-wave spiral arm. The Carina region  has kinematics typical
for a spiral arm inside the CR, whereas the Sagittarius region has
kinematics typical for its being outside the CR. However, their
kinematics contradicts their location: the Sagittarius region
($R=5.6$ kpc) is located closer to the Galactic center than the
Carina region ($R=6.5$ kpc). Thus, the simple trailing
density-wave spiral arm is not applicable to the Galaxy.

The  velocity $V_\theta$ is conspicuous only in the Carina,
Cygnus, and Perseus  regions. In the Carina region the  velocity
$V_\theta$ is directed in the sense of Galactic rotation
($V_\theta=+5$ km s$^{-1}$), while in the Perseus  ($V_\theta=-5$
km s$^{-1}$) and Cygnus ($V_\theta=-11$ km s$^{-1}$) regions  it
is in the opposite sense.

\subsection[]{Morphology and modelling of the outer pseudorings}

The essential characteristic of the galaxies with the outer rings
and pseudorings -- incomplete rings made up of spiral arms -- is
the presence of the bar (Buta 1995; Buta and Combes 1996). Two
main classes of the outer rings and pseudorings have been
identified: the $R_1$ rings ($R'_1$ pseudorings) elongated
perpendicular to the bar and the $R_2$ rings ($R'_2$ pseudorings)
elongated parallel to the bar. In addition, there is a combined
morphological type $R_1R_2'$ which shows elements of both classes.
The $R_2$ rings have elliptical shape, but the $R_1$ rings are
often ``dimpled'' near the bar ends. There is also a lot of outer
rings/pseudorings that cannot be classified into previous classes
because their morphological characteristics are unclear or the
inclination prevents detailed classification (Buta 1995; Buta and
Crocker 1991; Buta et al. 2007).

The outer rings are typically observed in early-type galaxies. For
the galaxies in the lower red-shift range the frequency of the
outer rings is found to be about 10\% of all types of spiral
galaxies. But for the early-type sample it increases to 20\% (Buta
and Combes 1996). A study by Buta (1995) shows the following
distribution among the main outer ring types: 18\% ($R_1$), 37\%
($R_1'$), under 1\% ($R_2$), 35\% ($R_2'$), and 9\% ($R_1R_2'$).
The small fraction of the complete $R_2$  rings may be due to the
selection effects -- they lack conspicuous features, for example,
"dimples", and so their definite classification can be difficult
due to orientation uncertainties.

The test particle simulations  (Schwarz 1981; Byrd et al. 1994;
Rautiainen and Salo 1999) and N-body simulations (Rautiainen and
Salo 2000) show that the outer pseudorings   are typically located
in the region of the Outer Lindblad Resonance (OLR) and    are
connected with two main families  of periodic orbits. Main
families of stable  periodic orbits are followed by most
non-periodic orbits after the introduction of the bar potential
(Contopoulos and Papayannopulos 1980). The $R_1$ rings are
supported by periodic orbits lying inside the OLR and elongated
perpendicular to the bar, while the $R_2$  rings are supported by
orbits situated outside the OLR and elongated along the bar.

As convincingly shown by Schwarz (1981) the pseudorings appear
before the pure rings. According to Schwarz (1981, 1984), the type
of an outer ring is determined by two factors: the bar's strength
and the initial distribution of gas particles. Under the stronger
bar forcing the pure $R_1$ ring no longer forms, instead,
particles move outside through the OLR and form the pseudoring
$R_2'$. However, this idea contradicts observations: the $R_2'$
pseudorings are more frequently observed in SAB galaxies than in
SB ones (Buta 1995).

Simulations demonstrate that such factors as the strong bar, a
large radius of the initial particle distribution, and a large
model time favor the formation of the $R_2$ component. Byrd et al.
(1994) find that the $R_1$ component appears quickly and the $R_2$
component forms slower.  All their models yield the $R_1'$
pseudoring at the earlier time and the $R_2'$  or the combined
$R_1R_2'$ pseudoring at the later time.

In some galaxies with the combined morphology the $R_1$ component
can be seen in infrared, but the $R_2'$ component is usually
prominent only in blue. Byrd et al. (1994) explain this fact by
the age difference between two components. N-body simulations
suggest another explanation. Rautiainen and Salo (2000) think that
at least some stellar $R_1'$ rings are not remnants of previous
ring-shaped star formation episode, but forms due to self-gravity
in the stellar subsystem.

Galactic disks in some simulations demonstrate the presence of the
slow modes near the radius of the outer ring (Rautiainen and Salo
1999, 2000). We like to emphasize that the present paper explains
the kinematics of young stars without invoking the slow modes.

\subsection[]{The  bar, rotation curve, and a prototype of the Galaxy}

Our Galaxy certainly has the bar. The gas kinematics in the
central region, infrared photometry, star counts, and other modern
tests confirm this fact (Weiner and Sellwood, 1999; Benjamin et
al. 2005; Englmaier and Gerhard 2006; Habing et al. 2006;
Cabrera-Lavers et al. 2007; and other papers). There is ample
evidence suggesting that the major axis of the bar is oriented in
the direction $\theta_b=15\textrm{--}45^\circ$ in such a way that
the end of the bar closest to the Sun lies in the first quadrant.
However, the angular speed of the  bar $\Omega_b$ and its length
are determined from observations badly. Some researchers believe
that the CR of the bar lies at the distance range
$R=3\textrm{--}4$ kpc (Englmaier and Gerhard 2006; Habing et al.
2006; and references therein), whereas others suggest that the
Galaxy has a longer bar with the major axis of $a=4\textrm{--}5$
kpc (Weiner and Sellwood, 1999; Benjamin et al. 2005;
Cabrera-Lavers et al. 2007; and references therein).

Analysis of orbits in  barred galaxies shows that a bar cannot
reach beyond its CR -- outside this radius the main family of
periodic orbits becomes oriented perpendicular to the bar, thus
unable to support it (Contopoulos and Papayannopulos 1980). This
sets an upper limit for the angular speed of the bar. However, the
lower limit is less clear. In general the CR is believed to lie
within 1.0 -- 1.4 $R_{bar}$, although higher values for
$R_{CR}/R_{bar}$ has also been suggested for some galaxies (for
example, Rautiainen et al. 2005).

For our study the more important thing is the location of the OLR
which is supposed to lie in the solar neighborhood (Kalnajs 1991;
and other papers). For a flat rotation curve, the location of the
OLR between the Sagittarius ($R=5.6$ kpc) and Perseus ($R=8.4$
kpc) regions corresponds to $\Omega_b$ lying within the limits
$\Omega_b=43\textrm{--}65$ km s$^{-1}$ kpc$^{-1}$ and the CR
within the range $R_{CR}=3.3\textrm{--}5.0$ kpc.

Observations don't give an unambiguous answer on the question
about the form of the Galactic rotation curve in the central
region, $R<3$ kpc. The line-of-sight velocities at the tangential
points indicate a peak ($R \approx 0.3$ kpc)  and a local minimum
($R\approx 3$ kpc) on the rotation curve, though the depth of the
minimum is less than 50 km s$^{-1}$ (Burton and Gordon 1978).
However, the apparent peak and minimum  can be due to the
perturbation of the circular velocities by the  bar (Englmaier and
Gerhard 2006; and other papers). In the outer region the Galactic
rotation curve is nearly flat. The line-of-sight velocities for
HII-regions, molecular clouds, and Cepheids suggest this idea
(Brand and Blitz 1993; Dambis et al. 1995; Russeil 2003; and other
papers). Our hypothesis is that the Galactic rotation curve is
nearly flat at the distance range $R=1-12$ kpc, though it can have
a small peak and a local minimum in the central region.

As will be shown below, the  kinematics  of young stars in the
Perseus  region  indicates the existence of the $R_2$ ring, while
the velocities in the Sagittarius region suggest the presence of
the $R_1$ ring in the Galaxy.  We suggest that the Galaxy has the
combined morphology $R_1R_2'$. Buta and Crocker (1991) regard the
galaxies ESO 509-98 and ESO 507-16 as typical examples of the
$R_1R_2'$ morphology.  Here are some other examples of galaxies
with  the $R_1R_2'$ morphology which can be also supposed as
possible prototypes of the Galaxy: ESO 245-1, NGC 1079, NGC 1211,
NGC 3081, NGC 5101, NGC 5701,  NGC 6782, and NGC 7098.

\section[]{Models}

\begin{figure*}
 \resizebox{\hsize}{!}{\includegraphics{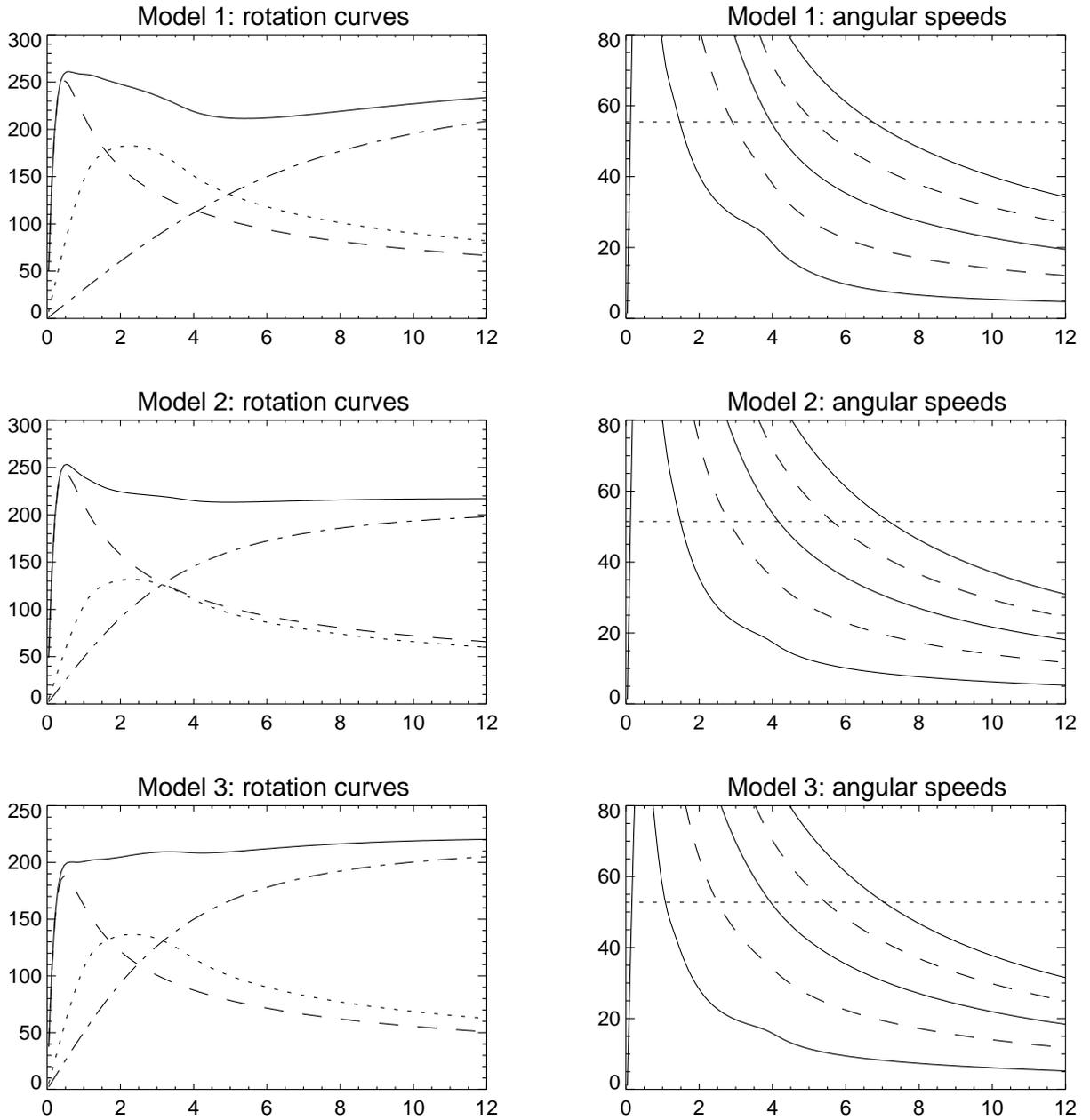}}
\caption{ Left panel: the rotation curves in models 1--3. The
continuous lines show the total curve (km s$^{-1}$), while the
dashed, dotted and dash-dotted lines show the contribution of the
bulge, bar, and halo, respectively. Right panel: the curves of the
angular speeds. The continuous lines indicate the angular speeds
$\Omega_0$ and $\Omega_0 \pm \kappa/2$, while the dashed lines --
$\Omega_0 \pm \kappa/4$ (km s$^{-1}$ kpc$^{-1}$). The straight
dotted line represents the angular speed of the bar $\Omega_{bar}$
in each case. The resonance distances are determined by its
intersections with the curves of angular speeds. Distances are
given in kpc.}
\end{figure*}

%--------- Table 2  ------------------
\begin{table}
\begin{center}
\small
 \caption{The potential components}\footnotesize
\begin{tabular}{lccc}
\\[-7pt]\hline\\[-7pt]
Model & 1 & 2& 3\\
\hline
bulge & & &\\
mass [$10^{10} \ M_\odot$]& 1.22 & 1.21 & 0.72\\
$r_b$ [kpc] & 0.31 & 0.33 & 0.34\\
\hline
halo & & &\\
$r_c$ [kpc] & 8.0 & 4.15 & 4.20\\
$v_{max} \ [\mathrm{km \ s^{-1}}]$ & 251.6 & 209.5 & 217.3\\
\hline
bar & & &\\
mass [$10^{10} \ M_\odot$]& 1.82 & 0.99 & 1.08\\
a,b [kpc] & 3.82, 1.20 & 3.98, 1.25 & 4.03, 1.26\\
\hline
\end{tabular}
\end{center}
\end{table}

The simulation program we use was developed by Salo (1991). It has
been  used in both self-gravitating simulations (Rautiainen and
Salo 1999) and models where the gravitational potential  has been
derived from infrared observations (for example, Rautiainen et al.
2005). Here we use analytical expressions for the potential
components, as was also done by Byrd et al. (1994) in simulations
with an earlier version of the same code.

We constructed a series of models which have essentially flat
rotation curve in the region corresponding to the solar
neighborhood. The galactic potential in all models consists of
three components: a bulge, halo, and a bar (Fig.~2). We have not
additionally included a disc component, because a flat rotation
curve can be achieved without it.

The bulge potential is a Plummer sphere (see, for example, Binney
and Tremaine 2008) which defines the slope of the rotation curve
in the inner region. The outer part is dominated by a halo whose
rotation curve is

\begin{equation}
 v^2(r)=v_{max}^2r^2/(r^2+r_c^2),
\end{equation}

\noindent where $v_{max}$ is the asymptotic maximum on the halo
rotation curve and $r_c$ is a core radius. This should not be
considered as a pure halo component, because, strictly speaking,
we do not make specific assumptions of the halo-disc mass ratio.
We are making two-dimensional simulations without self-gravity:
the particles are feeling the axisymmetric potential (essentially
the rotation curve and its slope, see Binney and Tremaine 2008)
and the non-axisymmetric bar perturbation. The disc in this case
can be thought to consist both of the bar and part of the halo.
This approach is common in studies of ring formation or orbits in
barred galaxies.  The gravitational effects of the rings and
spiral arms are omitted in this stage of study.

The bar is modelled as a Ferrers ellipsoid, whose density
distribution is

\begin{equation}
\rho = \left\{ \begin{array}{l}
\rho_0 (1-m^2)^n, \ m \leq 1\\
0, \ m > 1,\\
\end{array} \right.
\label{kalnajs}
\end{equation}

\noindent where $m$ equals $m^2=x^2/a^2+y^2/b^2$ and in our case
$n=1$. The Ferrers ellipsoid is often used in gasdynamical
simulations and in orbital analysis (for example, Athanassoula
1992; Romero-Gomez et al. 2007).

The  parameters of the previously mentioned gravitation potential
components were scaled so that the linear velocity of the rotation
curve at the solar distance equals observational one (215 km
s$^{-1}$). We achieved the best fit between the model and observed
velocities using small variations in the scale and in the solar
position angle. In particular, we suppose that the Sagittarius
region (5.6 kpc) is lying near the point D of the $R_1$ ring,
where the velocities $V_\theta$ equal zero (see section 3.2). In
simulation units the values of the bar's semi-axes, $a$ and  $b$,
are the same in all models, but the fine-tuning introduces small
differences in the length scales (Table 2). One time unit in our
simulations corresponds to the physical time of
$T=62\textrm{--}75$ Myr. The angular speed of the bar has the
value of $\Omega_{bar}=52\textrm{--}55$ km s$^{-1}$ kpc$^{-1}$
which corresponds to the bar rotation period of
$T=112\textrm{--}118$ Myr.

In our models, the non-axisymmetric perturbation is gradually
turned on during four bar rotation periods. This was done to avoid
possible transient effects. The initial stage assumes only small
deviations from the circular motion. One should especially note
that the bar mass is included in the models from the beginning:
initially only its $m=0$ component has a non-zero value.

The gas subsystem is modelled by 50 000 massless test particles
(the initial surface density is uniform in the occupied region)
that can collide with each other inelastically. Initially, the gas
disk is cold, its velocity dispersion corresponds to only few km
s$^{-1}$. The velocity dispersion rises during simulation, but
does not exceed the typical values seen in the galaxies. To detect
regions corresponding to OB-associations, we used a simple recipe
of the star formation: in each collision there is a small
probability (typically 0.1) that the gas cloud becomes a massless
``OB-association''. These associations have a limited lifetime
before returning as gas particles. With the adopted units of
lengths and velocities  it corresponds to about four million
years.

Our analysis includes  both gas particles and OB-particles. Though
their behavior  is quite similar, there are also some differences:
the collision-based star formation recipe favors regions where
orbits are crossing. Thus, the density of OB-particles is not
directly proportional to  the density of gas clouds.

Fig.~3 shows the distribution of OB-particles in  models 1--3 at
the different time steps.  The  evolution of models can be briefly
described as follows. When the bar is turned on a nuclear ring and
an inner ring  are quickly forming  near  the inner Lindblad
resonance and the inner 4/1-resonance, respectively (see, for
example, Buta and Combes 1996). These rings also quickly
disappear, only model 3 keeps a well-defined inner ring by the
moment $T=7.5$. As for the outer region, all models form an outer
pseudoring $R_2'$ by the time $T=7.5$ and nearly pure rings of
class $R_1R_2$ by the moment $T=15$. In its early evolution the
$R_2'$ pseudoring is not oriented strictly parallel to the bar,
its major axis is slightly leading with respect to the bar
(T=11.5, models 1 and 3), but by the time $T=15$ it  adjusts to
parallel or almost parallel alignment with the bar. In a long
simulation, the $R_1$ and $R_2$ rings acquire a more circular form
(T=25). In the case of the strong bar (model 1) the evolution is
faster in general.

The evolution in the outer region depends also on the initial gas
particle distribution. If the gas disc is very extended, an outer
pseudoring of class $R_2'$ can form almost as fast as the $R_1'$
component, but if its extent is small ($R_{cut}<R_{OLR}$), the
$R_2'$ component forms more slowly. A very small radius of the
initial particle distribution can  delay the appearance of the
outer pseudorings and even inhibit the formation of an $R_2'$
pseudoring.

\begin{figure*}
\resizebox{0.9\hsize}{!}{\includegraphics{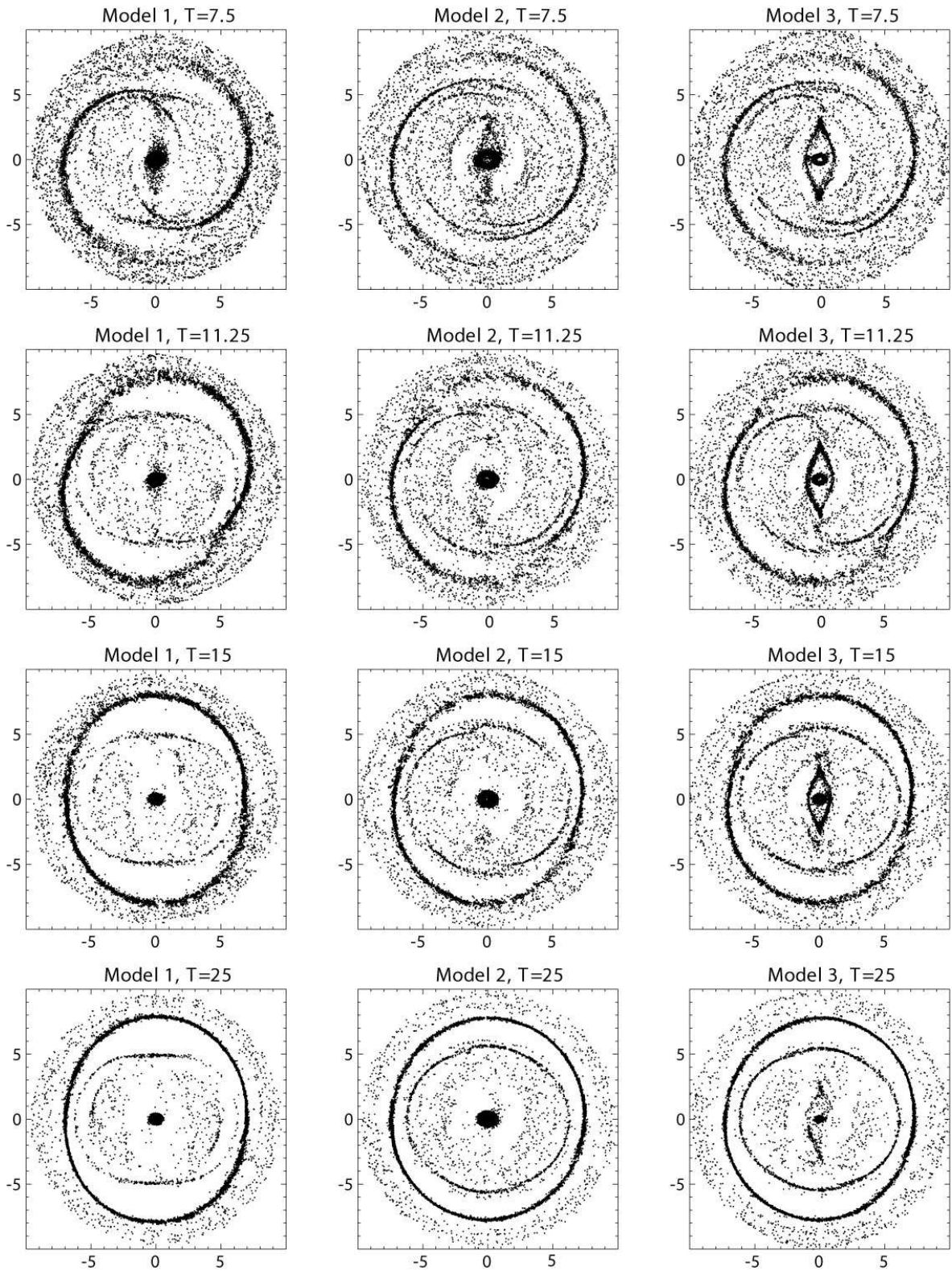}} \caption{The
distribution of OB-particles  in models 1--3 at $T=7.5$, $11.25$,
$15.0$, and $25.0$.}
\end{figure*}

\section[]{Kinematics of the outer rings and pseudorings}

\subsection[]{The resonance kinematics}

%--------- Table 3  ------------------
\begin{table}
\begin{center}
 \caption{Amplitudes of the velocity oscillations
  at  the distance of $R=7.38$ kpc}\small

\begin{tabular}{lccc}
\\[-7pt]\hline\\[-7pt]
   Model & 1 & 2 & 3  \\
  \hline
  T=7.5 & & & \\
  $f_R$, $f_\theta$, km s$^{-1}$  & 24, 16 & 22, 12 & 24, 13  \\
  \hline
  T=15 & & & \\
  $f_R$, $f_\theta$, km s$^{-1}$  & 31, 15 & 23, 12 & 25, 12 \\
  \hline
  T=25 & & & \\
  $f_R$, $f_\theta$, km s$^{-1}$  & 25, 12 & 12, 6 & 15, 8 \\
  \hline
\end{tabular}
\end{center}
\end{table}

The modelling suggests that the kinematics of the outer rings and
pseudorings is determined by  two processes: the resonance tuning
and the gas outflow.  The resonance kinematics is clearly observed
in the pure rings, while the kinematics of the gas outflow is
manifested itself in the pseudorings.

Fig.~4 exhibits the distribution of gas particles and OB-particles
with the positive and negative residual velocities in à projection
onto the radial and azimuthal directions at the time moment $T=15$
($\sim 8$ bar rotation periods). Model 1 was chosen for
illustration only, other models display a very similar velocity
distribution. The residual velocities were calculated as
differences between the model velocities and the velocities due to
the rotation curve. The  alternation of the quadrants with the
negative and positive residual velocities is clearly observed in
the pure rings. Besides, particles located at the same azimuthal
angle $\theta$ but forming either the $R_2$ or the $R_1$ ring have
the opposite residual velocities. The oscillations of the
azimuthal velocities, $V_\theta$, are shifted by the angle
$\theta\approx \pi/4$ with respect to the radial-velocity
oscillations.

\begin{figure*}
\resizebox{\hsize}{!}{\includegraphics{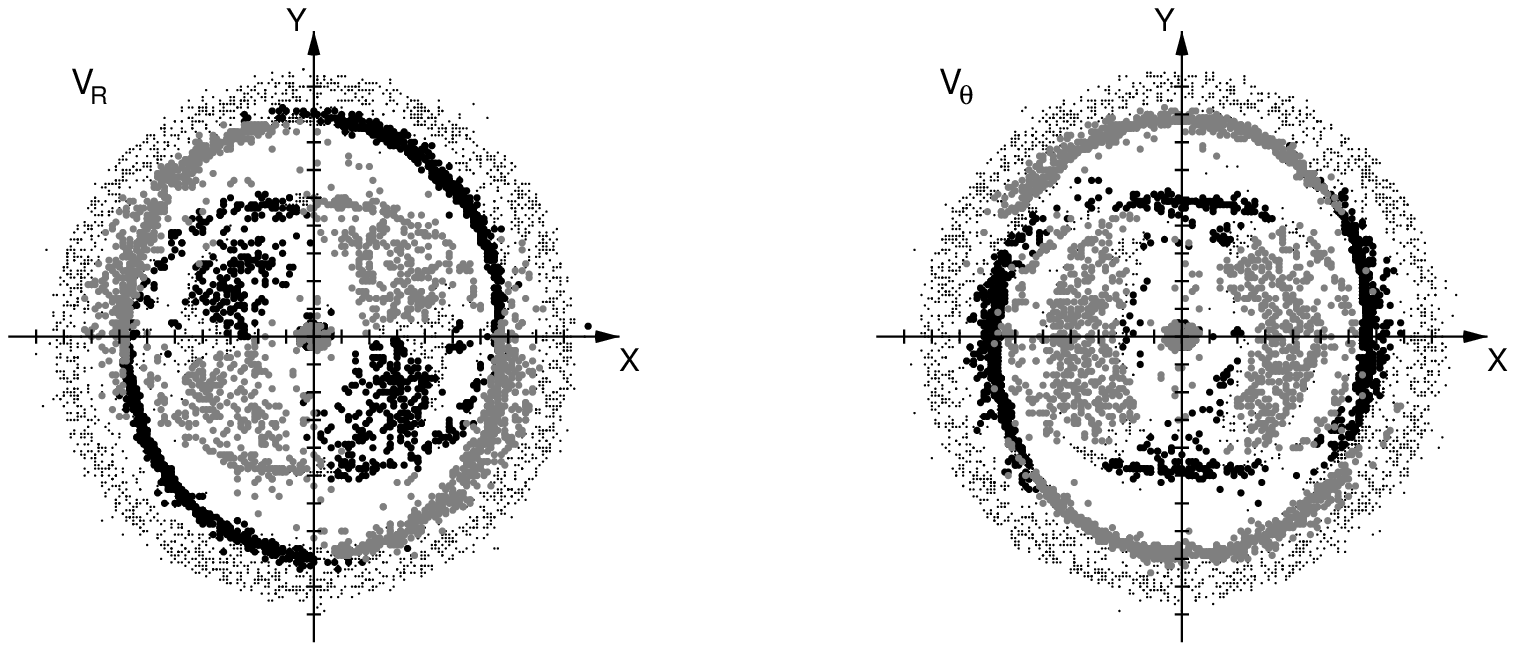}} \caption{The
distribution of particles (gas+OB) with the negative and positive
residual velocities  in model 1 at the moment $T=15$. The bar is
oriented along the Y-axis, the galaxy rotates clockwise, a
division on the $X$- and $Y$-axis corresponds to 1 kpc. The left
panel represents the radial  velocities, while the right one --
the azimuthal ones. Only 20\% of particles are shown. Particles
with the positive residual velocities ($V_R>5$ km s$^{-1}$ or
$V_\theta>5$ km s$^{-1}$) are indicated by black circles, while
particles with the negative ones ($V_R<-5$ km s$^{-1}$ or
$V_\theta<-5$ km s$^{-1}$) --  by  grey circles. Particles with
the velocities close to zero ($-5<V_R<+5$ km s$^{-1}$ or
$-5<V_\theta<+5 $ km s$^{-1}$) are shown by points.}
\end{figure*}

To  study  oscillations  of the residual velocities we selected
particles (gas+OB) located in the narrow annulus of
$R=7.38\pm0.16$ kpc in different models at different moments.
Fig.~5 shows the velocity oscillations of selected particles in
model 2 at $T=15$. The position of the selected particles
corresponds to the average radius of the $R_2$ ring. The velocity
profiles made at the same radius but for different models and
moments resemble each other, the main difference is seen in
velocity amplitudes. Table 3 represents the amplitudes of velocity
oscillations, $f_R$ and $f_\theta$, calculated for selected
particles in different models at different moments. The highest
velocity amplitudes are observed in model 1, while the amplitudes
in models 2 and 3 are lower. In each model the lowest velocity
amplitudes are observed at $T=25$.

\begin{figure}
\resizebox{\hsize}{!}{\includegraphics{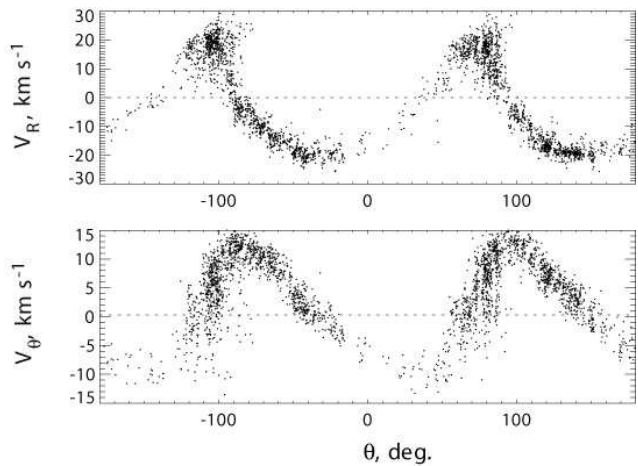}} \caption{The
oscillations of the velocities, $V_R$ and $V_\theta$, of particles
(gas+OB) located in the narrow annulus $R=7.38\pm0.16$ kpc in
model 2 at $T=15$.}
\end{figure}

We combined the samples of gas particles and OB-particles, because
their kinematics is very similar. That is not surprising, because
"OB-associations" in our models have the velocities of their
parent gas clouds.

\subsection[]{The orbital kinematics in the resonance region}

The resonance between  the epicyclic  and orbital motions adjusts
the epicyclic motions of gas clouds in accordance with orbital
rotation. That creates  systematical non-circular motions of gas
clouds whose direction  depends on the position angle of a point
with respect to the bar major axis and on the class of the outer
ring.

Fig.~6 shows the directions of the residual velocities at the
different points of the outer pseudorings. It schematically
represents the bar and two main families of periodic orbits. The
galaxy rotates {\it clockwise}. Motions are considered in the
reference frame rotating with the angular speed of the bar. In
this frame stars located near the OLR rotate with the speed
$(\Omega-\Omega_b)$ {\it counterclockwise}.

Let us consider the family of periodic orbits elongated along the
bar and located outside the OLR (Fig.~6). It is represented by the
orbit passing through the points 1-2-3-..-8. This family is the
backbone of the $R_2$  ring. The points 1 and 5 are the
peri-centers of the orbit, but the points 3 and 7 are the
apo-centers. The epicycles drawing along the border of the figure
demonstrate the position of a particle on the epicyclic orbit at
the points 1--8. The arrows show the additional velocities caused
by the epicyclic motion which can be regarded as the residual
velocities. Let us consider the projection of the residual
velocities onto the radial and azimuthal directions. At the points
2 and 6 the velocity $V_R$ is directed away from the galactic
center and achieves its extremal positive value, while at the
points 4 and 8 it is directed toward the galactic center and
exhibits its extremal negative value. Stars  have the negative
radial velocities on the orbital segments 3-5 and 7-1 which are
marked by dashed lines. Azimuthal residual velocity, $V_{\theta}$,
equals zero at the points 2, 4, 6, and 8; it achieves its extremal
positive amount at the points 1 and 5 and its extremal negative
value at the points 3 and 7.

Another family of periodic orbits oriented perpendicular to the
bar is represented by the orbit passing through the points
A-B-C-..-H (Fig.~6). The "dimples" near the points C and G appear
due to a relatively large size of the epicycle in comparison with
the orbital radius. This family is the backbone of the $R_1$ ring.
The radial velocity, $V_R$, achieves its extremal positive value
at the points D and H and the extremal negative one at the points
B and F.  Points B, D, F, and H are located not at the middles of
the corresponding segments but a bit closer to the bar's ends. The
azimuthal velocity, $V_{\theta}$, is zero at the points B, D, F,
and H; it achieves its extremal positive value at the points C and
G and the extremal negative one at the points A and E.

The directions of the residual velocities in models 1--3 at the
moment $T=15$ (Fig.~4) are in a good agreement with the resonance
kinematics shown in Fig.~6. Probably, the kinematics of the pure
rings (T=15, T=25) is determined by the resonant orbits only.

In the epicyclic approximation the elongation of the outer rings
can be explained through the subtraction  and addition of the size
of the epicycle to the average radius of the ring. In this case
the outer rings must demonstrate the following tendency: the
smaller the elongation of the ring, the  smaller the  residual
velocities. Models 1--3 confirm this dependence. The outer rings
become more round by the time $T=25$ (Fig.~3), that is accompanied
with the decreasing velocity amplitudes (Table 3).

\begin{figure} \resizebox{\hsize}{!}{\includegraphics{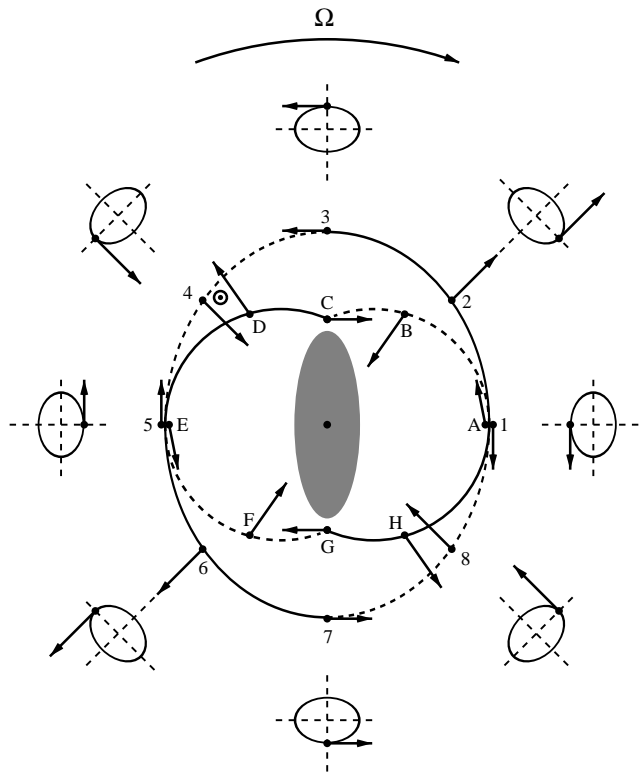}}
\caption{Orbital kinematics in the resonance region. It shows the
bar and two main families of periodic orbits in the region of the
OLR. The galaxy rotates clockwise. Motions are considered in the
reference frame rotating with the speed of the bar. The orbit
going through the points 1-2-3-4-5-6-7-8 represents the family of
periodic orbits elongated along the bar. The orbit A-B-C-D-E-F-G-H
denotes the family of periodic orbits elongated perpendicular to
the bar. The epicycles drawing on the border of the picture show
the position of a particle on the epicyclic orbit at the points
1--8. The vectors show the directions of the additional (residual)
velocities due to the epicyclic motion. The dashed lines indicate
the orbital segments with the negative radial velocities, while
the solid lines --  those with the positive velocities $V_R$. The
likely position of the Sun is shown by the specific symbol.}
\end{figure}

\subsection[]{Kinematics of the gas outflow}

The main kinematical feature of the outer pseudorings is the
preponderance of  positive radial velocities against  negative
ones. All models considered demonstrate the gas outflow  at early
moments of their evolution. Fig.~7 represents the distribution of
particles (gas+OB) with the positive and negative residual
velocities in  model 1  at $T=7.5$ ($\sim 4$ bar rotation
periods), other models have a very similar velocity distribution.
Left panel of Fig.~7 clearly shows that particles with the
positive radial velocities concentrate to the spiral arms. Right
panel of Fig.~7 demonstrates another kinematical feature of the
pseudorings: particles located in the inner parts of the spiral
arms ($R<R_{OLR}$) have only positive azimuthal velocities
$V_\theta$, while those forming the outer parts of the spiral arms
($R>R_{OLR}$) have only negative ones.

\begin{figure*}
\resizebox{\hsize}{!}{\includegraphics{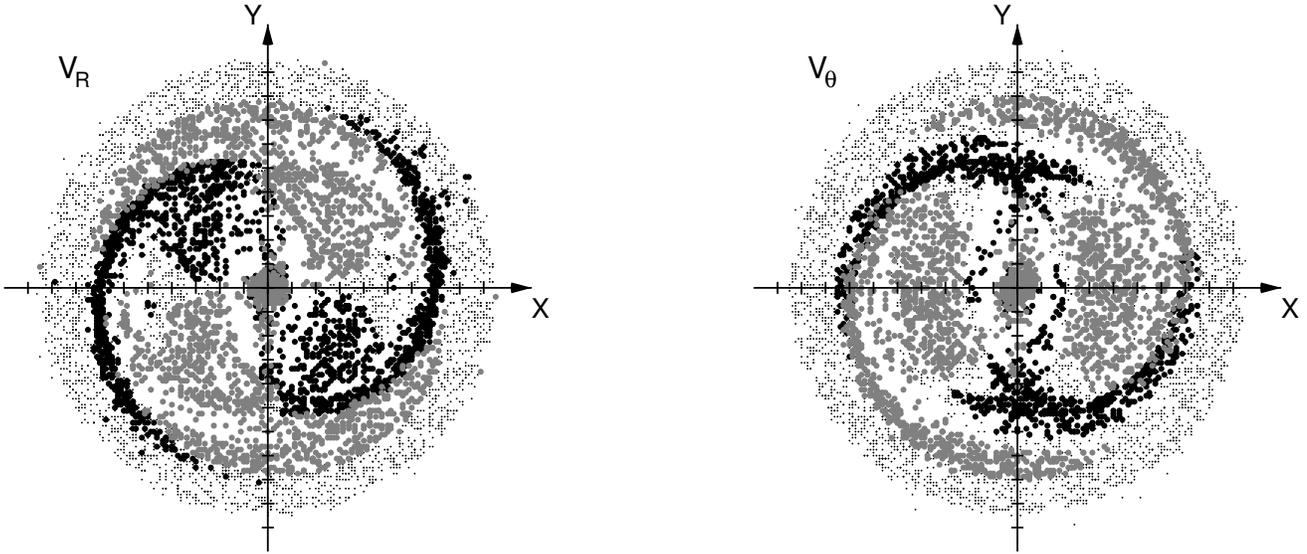}} \caption{The
distribution of gas+OB particles with the negative and positive
residual velocities in model 1 at T=7.5. Particles with the
positive residual velocities are indicated by black circles, while
particles with the negative ones -- by grey circles. Particles
with the residual velocities close to zero are shown by points.
For more details see the caption to Fig.~4.}
\end{figure*}

Schwarz (1981, 1984) finds that the gas outflow to the periphery
plays an important role in the formation of the outer rings.
Fig.~8 shows the profiles of the surface density of particles
(gas+OB) in  models 1--3 at different moments. At the initial
moment the surface density  is constant in the interval
$R=1\textrm{--}10$ kpc in all models. At the moment $T=15$ all
density profiles have a well-defined maximum corresponding to the
location of the $R_2$  ring ($R=7\textrm{--}8$ kpc). It grows
faster in model 1 than in models 2 and 3 what indicates  the more
intense gas outflow in model 1. In all models the maximum
corresponding to to the $R_1$ ring is lying in the vicinity of the
outer 4/1 resonance, while the maximum corresponding to the $R_2$
ring is shifted $\sim 0.5$ kpc outside the OLR.

\begin{figure}
\resizebox{\hsize}{!}{\includegraphics{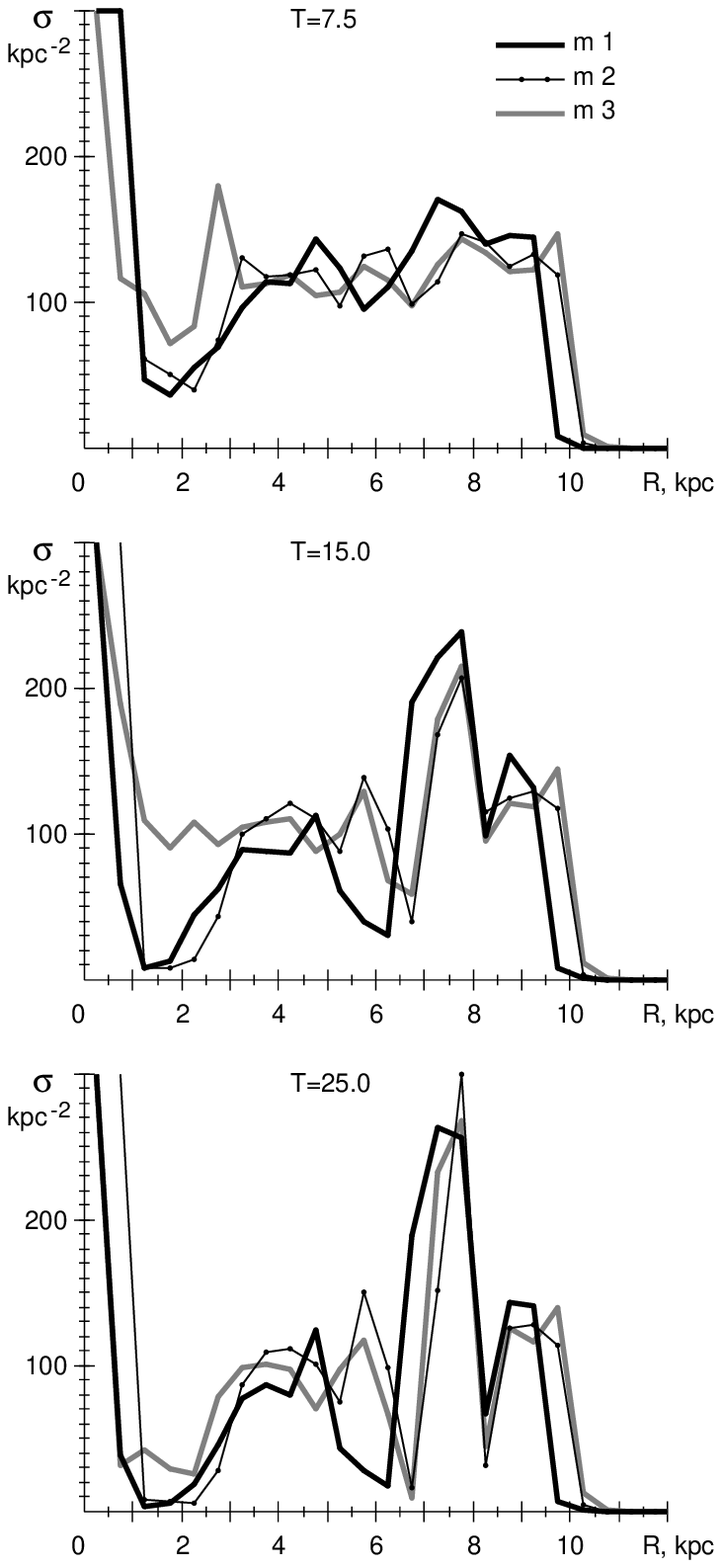}}
\caption{Profiles of the surface density of particles (gas+OB)  in
models 1--3 at  different moments. At the initial moment the
surface density is constant in the interval $R=1-10$ kpc.}
\end{figure}

\section{Comparison with the Galaxy}

\subsection{Positions of the stellar-gas complexes with respect to the outer
rings}

Let us suppose that the kinematics of young stars in the solar
neighborhood is determined by their position with respect to the
outer rings. Observations suggest that  the Sun goes
$\theta_b=15\textrm{--}45^\circ$ behind the bar elongation, so all
complexes studied must be located near the segments 3--5 and C--E
of the outer rings (Fig.~6). The Perseus, Cygnus, and Carina
regions having the negative velocities $V_R$ must belong to the
ring/pseudoring $R_2$, but the Sagittarius region with the
positive velocities $V_R$ -- to the ring/pseudoring $R_1$.

There is some problem with the Perseus ($R=8.4$ kpc) and Carina
($R=6.5$ kpc) regions: they  have very different Galactocentric
distances $R$ and  they both cannot accurately lie on the same
ring. Otherwise the ring $R_2$ must be highly elongated: the ratio
$q$ of the minor and major axis  must be of $q\approx0.6$.
However, the observations suggest that the rings/pseudorings $R_2$
are not so much elongated: their average axis ratio is $q=0.9$
(Buta and Combes 1996). So the Perseus region must lie a bit
outside the $R_2$  ring, while the Carina region -- a bit inside
it.

The kinematics of the Sagittarius region is very important for the
definition of a type of the Galactic morphology.  The  nearly zero
residual azimuthal velocities, $V_{\theta}=-1$ km s$^{-1}$, here
suggest the presence of the $R_1$ ring in the Galaxy but not of
the spiral arms. Particles located in the inner parts of the
spiral arms have only positive velocities $V_\theta$ (Fig.~7,
right panel) and  cannot reproduce $V_{\theta}=-1$ km s$^{-1}$.
Thus, the Galaxy must also include the $R_1$ ring.

\subsection{Comparison between models and observations}

\begin{figure*}
\resizebox{\hsize}{!}{\includegraphics{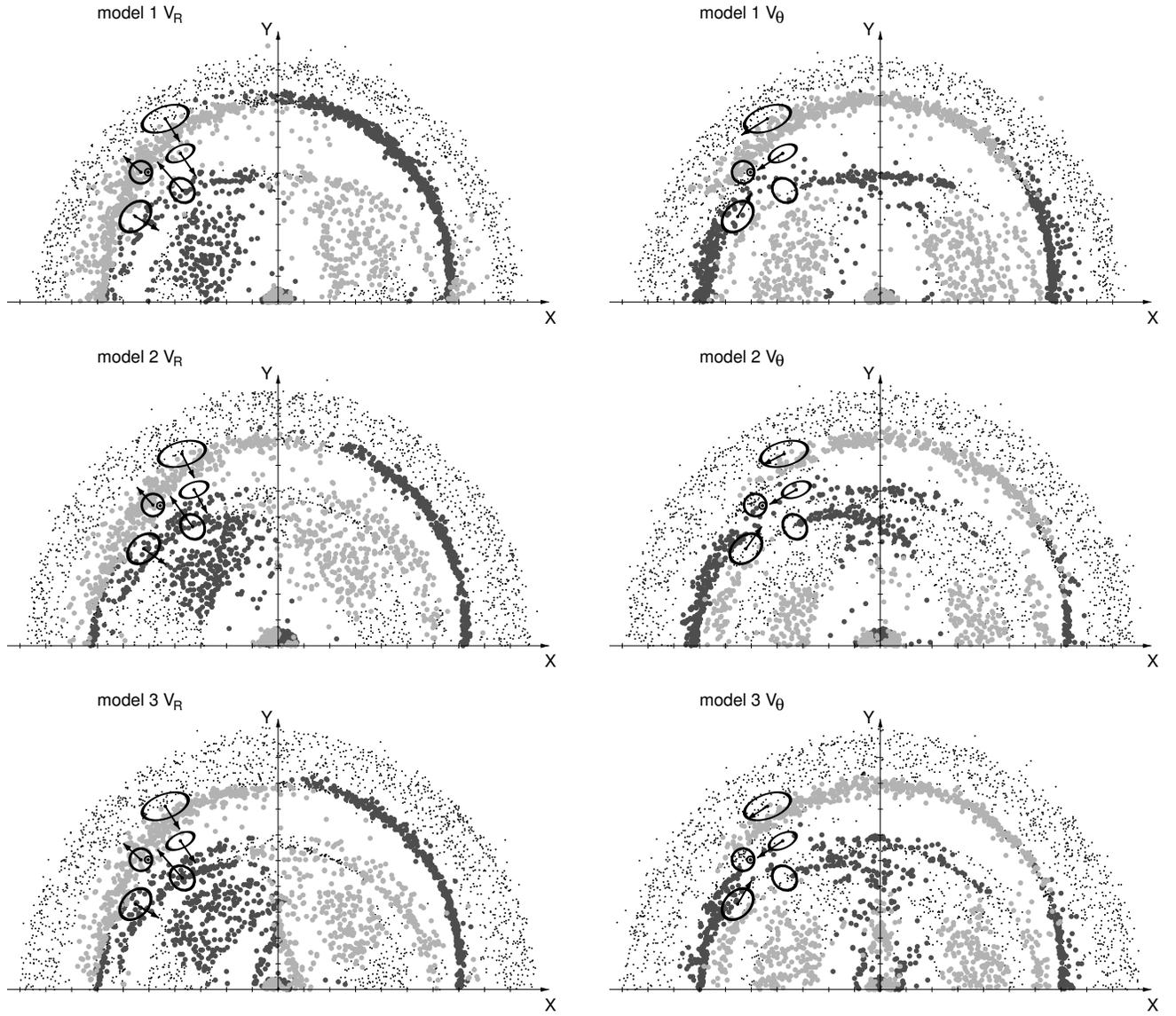}} \caption{The
boundaries of the stellar-gas complexes overlaid above the
distribution of particles with the positive and negative residual
velocities in models 1--3 at the moment $T=15$. The velocity
vectors represent the observed residual velocities. Particles with
the positive residual velocities ($V_R>5$  km s$^{-1}$ or
$V_\theta>5$ km s$^{-1}$) are indicated by black-grey circles,
while particles with the negative ones ($V_R<-5$ km s$^{-1}$  or
$V_\theta<-5$ km s$^{-1}$) -- by light-grey circles. Particles
with the residual velocities close to zero are shown by points.
For more details see the caption to Fig.~4.}
\end{figure*}

%--------- Table 4  ------------------
\begin{table}
\begin{center}
\caption{Model 1. Average residual velocities}\footnotesize
\begin{tabular}{lccccc}
  \hline
   Region & $V_R$ & $\sigma_R$ & $V_\theta$ &  $\sigma_\theta$ & n \\
          & km s$^{-1}$ & km s$^{-1}$ & km s$^{-1}$ &  km s$^{-1}$ &  \\
\hline
\multicolumn{6}{l}{T=7.5  } \\
Sagittarius &     21 &    10 &     10 &    12 &  164 \\
Carina      &     18 &    14 &      8 &    11 &  322 \\
Cygnus      &    -23 &     3 &      7 &     2 &   33 \\
Local System&    -15 &     4 &      3 &     2 &   63 \\
Perseus     &     -4 &     1 &     -1 &     2 &  191 \\
\hline
\multicolumn{6}{l}{T=15.0 } \\
Sagittarius &     24 &     6 &      0 &     5 &   41 \\
Carina      &     -4 &    20 &     10 &     9 &   91 \\
Cygnus      &    -24 &     4 &      1 &     3 &   18 \\
Local System&    -24 &     3 &     -5 &     3 &  159 \\
Perseus     &     -4 &     5 &     -5 &     7 &  207 \\
\hline
\end{tabular}
\end{center}
\end{table}

%--------- Table 5  ------------------
\begin{table}
\begin{center}
\caption{Model 2. Average residual velocities} \footnotesize
  \begin{tabular}{lccccc}
  \hline
   Region & $V_R$ & $\sigma_R$ & $V_\theta$ &  $\sigma_\theta$ & n \\
          & km s$^{-1}$  & km s$^{-1}$ & km s$^{-1}$ & km s$^{-1}$ &  \\
\hline
\multicolumn{6}{l}{T=7.5  } \\
Sagittarius &     11 &     3 &      0 &     3 &   87 \\
Carina      &     12 &     4 &      0 &     7 &  183 \\
Cygnus      &    -16 &     2 &     11 &     2 &   49 \\
Local System&    -13 &     2 &      6 &     4 &   50 \\
Perseus     &     -4 &     2 &     -2 &     2 &  197 \\
\hline
\multicolumn{6}{l}{T=15.0 } \\
Sagittarius &     10 &     3 &      2 &     3 &   98 \\
Carina      &      9 &     6 &      1 &     8 &  149 \\
Cygnus      &    -16 &     1 &      8 &     3 &    5 \\
Local System&    -20 &     2 &      1 &     3 &  127 \\
Perseus     &     -7 &     6 &     -5 &     4 &  187 \\
\hline
\end{tabular}
\end{center}
\end{table}

%--------- Table 6  ------------------
\begin{table}
\begin{center}
\caption{Model 3. Average residual velocities}\footnotesize
\begin{tabular}{lccccc}
  \hline
  \small
   Region & $V_R$ & $\sigma_R$ & $V_\theta$ &  $\sigma_\theta$ & n \\
          & km s$^{-1}$ & km s$^{-1}$ & km s$^{-1}$ &  km s$^{-1}$ &  \\
  \hline
\multicolumn{6}{l}{T=7.5  } \\
Sagittarius &     13 &     4 &     -1 &     4 &  105 \\
Carina      &     15 &     6 &      1 &     9 &  246 \\
Cygnus      &    -17 &     2 &     10 &     2 &   31 \\
Local System&    -13 &     2 &      6 &     3 &   43 \\
Perseus     &     -4 &     1 &     -2 &     1 &  179 \\
\hline
\multicolumn{6}{l}{T=15.0 } \\
Sagittarius &     13 &     4 &      1 &     4 &  106 \\
Carina      &      7 &     9 &      2 &    11 &  138 \\
Cygnus      &    -18 &     5 &      6 &     2 &   12 \\
Local System&    -21 &     2 &      1 &     3 &  144 \\
Perseus     &     -8 &     7 &     -6 &     5 &  191 \\
\hline
\end{tabular}
\end{center}
\end{table}

%--------- Table 7  ------------------
\begin{table*}
\begin{center}
\caption{Comparison between models and observations}\small
  \begin{tabular}{lcc|cr}
\\[-7pt]\hline \\[-7pt]
   Region & $V_{R\mbox{ mod}}$  & $V_{R\mbox{ obs}}$ & $V_{\theta\mbox{ mod}}$ &
   $V_{\theta\mbox{ obs}}$ \\[3 pt]
            & km s$^{-1}$ & km s$^{-1}$ & km s$^{-1}$ &  km s$^{-1}$   \\
 \hline
Sagittarius &$(+10, +24)$ &  $+11\pm3$ & $(-1,+2)$   & $-1\pm2$ \\
Carina      &$(-4, +9)$   &  $-6\pm2$  & $(+1, +10)$  & $+5\pm3$ \\
Cygnus      &$(-24, -16)$ &  $-7\pm3$  & $(+1, +8)$  & $-11\pm2$ \\
Local System&$(-24, -20)$ &  $+5\pm3$  & $(-5, +1)$  &  $0\pm3$ \\
Perseus     &$(-4, -7)$   &  $-7\pm2$  & $(-5, -6)$ & $-5\pm2$ \\
\hline
\end{tabular}
\end{center}
\end{table*}

To compare models with observations we chose the moment $T=15$
when the resonance kinematics is best-defined.  Fig.~9 shows the
boundaries of the stellar-gas complexes overlaid above the
distribution of particles with the positive and negative residual
velocities at the moment $T=15$. The solar position angle is
supposed to be $\theta_b=45^\circ$ in models 1 and 3 and
$\theta_b=40^\circ$ in model 2 (explanation will be given below).
Tables 4--6 exhibit the average residual velocities, $V_R$ and
$V_\theta$, of  particles located within the boundaries of the
stellar-gas complexes in models 1--3 at the moments $T=7.5$ and
$15.0$. They also contain the velocity dispersions, $\sigma_R$ and
$\sigma_\theta$, and the number $n$ of particles (gas+OB) which
appear to fall within the complex at a certain time moment.

Table 7 represents  the model and observed  residual velocities
for each region. The interval of the model velocities demonstrates
the minimal and maximal velocity obtained in models 1--3 at
$T=15$.

The Perseus region ($R=8.4$ kpc) has the negative values of the
average velocities, $V_R$ and $V_\theta$, in all models (Table 7).
Though the amplitude of the radial velocities at the average
distance of the $R_2$  ring ($R=7.4$ kpc) is  fairly large,
$f_R=23\textrm{--}31$ km s$^{-1}$ (Table 3, $T=15$), its value
drops to $f_R=4\textrm{--}7$ km s$^{-1}$ at the distance of
$R=8.4$ kpc. The decrease in the velocity amplitude is due to the
fact that the Perseus region is situated a bit outside the $R_2$
ring and has less elongated orbit. Altogether, our models
reproduce well the kinematics in the Perseus region: the
difference between the model and observed velocities don't exceed
3 km s$^{-1}$.

The Sagittarius region ($R=5.6$ kpc) lies near the point D of the
$R_1$ ring due to the proper scaling and choice of the solar
position angle (Fig.~6). Particles located near the point D have
the positive  velocity $V_R$ and nearly zero  velocity $V_\theta$.
Model 1 yields a very large value of the radial velocity, $V_R=24$
km s$^{-1}$, but models 2 and 3 create the velocities in a
reasonable range, $V_R=10\textrm{--}13$ km s$^{-1}$. In models 2
and 3 the difference between the model and observed velocities
don't exceed 3 km s$^{-1}$. Interestingly, that at the moment
$T=7.5$  model 1 produces the large positive azimuthal velocity in
the Sagittarius region ($V_\theta=+10$ km s$^{-1}$), but by the
moment $T=15$ its value drops to $V_\theta=0$ km s$^{-1}$ (Table
4). Probably, the large velocity $V_\theta$ in model 1 is due to
the intense gas outflow at $T=7.5$. Models 2 and 3 do not show
such feature (Tables 5-6).

The Carina region ($R=6.5$ kpc) consists of two groups of
particles with different velocities (Fig.~9).  One group belongs
to the ring $R_1$, but another -- to the ring $R_2$. The mixture
of two streams produces  the larger values of the  velocity
dispersions, $\sigma_R$ and $\sigma_\theta$, here in comparison
with the other regions (Tables 4-6, $T=15$). The velocities of
particles in the ring $R_2$ are consistent with observations, so
the increase of their relative number in the Carina region brings
approaching between the model and observed velocities. All models
reproduce the direction of the observed azimuthal residual
velocity, but only model 1 reproduces the direction of the radial
one. Model 1 has more elongated ring $R_2$, what causes more
particles of the ring $R_2$ to fall within the boundaries of the
Carina region. Note that a small rescaling of models can
significantly change the average velocity of particles in the
Carina region.

The Cygnus region ($R=6.9$ kpc)  is located between two outer
rings where only a few  particles are located. Although the
direction of the model radial velocity agrees with observations,
its absolute value is too high. As for the azimuthal velocities,
none of our models can reproduce the observed negative azimuthal
velocity here (Table 7).

The Local System ($R=7.4$ kpc) lies in the vicinity of the $R_2$
ring in all models. It contains more particles at the moment
$T=15$ than at $T=7.5$. In the context of morphology the moment
$T=7.5$ is more suitable for a comparison with observations,
because at this moment the break in the $R_2$  ring is fairly wide
yet (Fig.~3). Our models reproduce well the nearly zero azimuthal
velocity, but they cannot create the observed positive radial
velocity here (Table 7).

\begin{figure}
\resizebox{0.9\hsize}{!}{\includegraphics{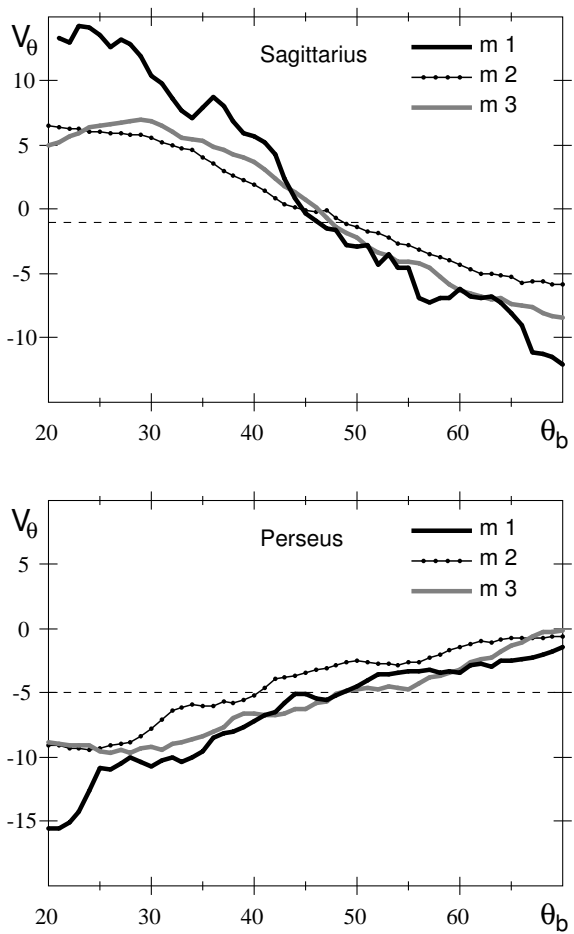}} \centering
\caption{The dependence of the model azimuthal velocity $V_\theta$
(km s$^{-1}$) in the Sagittarius and Perseus regions ($T=15$,
models 1--3) on the solar position angle $\theta_b$ (deg.)}
\end{figure}

Thus, models 2 and 3 reproduce well the values and directions of
the residual velocities  in the Perseus and Sagittarius regions,
but model 1 reproduces the directions  of the residual velocities
in the Perseus, Sagittarius, and Carina regions. Probably, the
kinematics of young stars in the Perseus, Sagittarius, and Carina
regions is  determined  mostly by the resonance, while  other
processes are dominant in the Cygnus region and in the Local
System.

The  value of the  model azimuthal velocity $V_\theta$ in the
Sagittarius and Perseus regions is especially sensitive to the
choice of the solar position angle $\theta_b$. An increase of the
angle $\theta_b$ causes a decrease of the velocity $V_\theta$ in
the Sagittarius region and an opposite effect in the Perseus one.
Fig.~10 exhibits the variations of the average velocity $V_\theta$
in the Sagittarius and Perseus regions with the change of the
angle $\theta_b$. It is seen that all models reproduce the value
of $V_\theta=-1$ km s$^{-1}$ in the Sagittarius region under
$\theta_b=45\textrm{--}48^\circ$. But in this range model 2
creates the small absolute values of the velocity $V_\theta$ in
the Perseus region, $|V_\theta|=3$ km s$^{-1}$. To get the larger
value of $|V_\theta|$ in Perseus we adopted $\theta_b=40^\circ$
for model 2.

\section{Conclusions}

The kinematics of the outer rings and pseudorings is determined by
two processes: the resonance tuning and the gas outflow.  The
resonance kinematics is clearly observed in the pure rings, while
the kinematics of the gas outflow is manifested itself in the
pseudorings. The resonance between the epicyclic and orbital
motion in the reference frame rotating with the bar speed adjusts
the epicyclic motions of particles in accordance with the bar
rotation. This adjustment creates the systematical non-circular
motions, whose direction depends on the position angle of a point
with respect to the bar elongation and on the class of the outer
ring.

Models of the Galaxy with the $R_1R_2'$ pseudoring reproduce well
the radial and azimuthal components of the residual velocities of
OB-associations in the Perseus and Sagittarius regions: the
difference between the model and observed velocities does not
exceed 3 km s$^{-1}$ (models 2 and 3, $T=15$). The kinematics  in
the Perseus region indicates the presence of the $R_2$ ring in the
Galaxy, while the velocities in the Sagittarius region suggest the
existence of the $R_1$ ring. The azimuthal velocities in the
Sagittarius region accurately defines the solar position angle
with respect to the bar elongation, $\theta_b=45\pm5^\circ$.
Besides, model 1 reproduces the directions of the radial and
azimuthal residual velocities in the Perseus, Sagittarius, and
Carina regions. Probably, the kinematics of young stars in the
Perseus, Sagittarius, and Carina regions is determined mostly by
the resonant orbits.

Our models have nearly flat rotation curve. The ring $R_1R_2'$ is
forming after several bar rotation periods. The $R_1$ ring lies in
the vicinity of the outer 4/1-resonance, while the $R_2'$
pseudoring is shifted $\sim 0.5$ kpc outside the OLR. We found
that the mixed $R_1R_2'$ morphology appears when the bar is not
very strong: in the case of a strong bar the $R_1$ ring
disappears. The strong bar and the initial  distribution of
particles in the disk of a large radius accelerate the formation
of the $R_2$  rings.

The major semi-axis of the bar  in our models has  the value of
$a=3.8\textrm{--}4.0$ kpc which amounts 55\% of the solar
Galactocentric distance, $R_0=7.1$ kpc, adopted here. The
relatively large size of the bar and the large value of the solar
position angle, $\theta_b=45\pm5^\circ$,  agree well with studies
in which authors find the presence of a long bar in the Galaxy
(Weiner and Sellwood, 1999; Benjamin et al. 2005; Cabrera-Lavers
et al. 2007; and other papers).

The model of the Galaxy with the $R_1R_2'$ pseudoring  can explain
some large-scale morphological features of the Galactic spiral
structure. The so-called Carina spiral arm (Russeil 2003) falls
nicely onto  the segment of the $R_2$ ring. Note that two outer
rings, which are stretched  perpendicular to each other, can look
like a 4-armed spiral pattern, especially if the ascending parts
of the rings are brighter than the descending ones. But this
spiral pattern appears not as a result of  the spiral perturbation
of the disk potential, but due to the existence of specific orbits
in the barred galaxies.

\acknowledgements

We wish to thank Heikki Salo for providing his simulation program.

This work  was partly supported by the Russian Foundation for
Basic Research (project no.~06\mbox{-}02\mbox{-}16077) and the
Council for the Program of Support for Leading Scientific Schools
(project no. NSh-433.2008.2).

\end{document}